\documentclass[
aps,%
11pt,%
final,%
notitlepage,%
oneside,%
onecolumn,%
nobibnotes,%
nofootinbib,%
superscriptaddress,%
noshowpacs,%
centertags]%
{revtex4}

\begin{document}
\selectlanguage{english}

\title{Galaxy cluster A2142: halo boundary, \glqq{red sequence}\grqq, properties
of galaxies according to SDSS}

\author{\firstname{F.G.}~\surname{Kopylova,}}
\affiliation{\saoname}

\author{\firstname{A.I.}~\surname{Kopylov}}
\affiliation{\saoname}

\onecolumngrid
\begin{center}
{\scriptsize
Original Russian Text @ F.G.~Kopylova, A.I.~Kopylov,
published in Astrofizicheskii Byulleten, 2022,\\
Vol.77, No.1, pp.24-33}
\end{center}

\begin{abstract}
\noindent

The paper presents the  results of studying the dynamics of galaxies,
properties
of early-type galaxies, properties of galaxies with the quenched star
formation (QGs) in the A 2142 based on the archival data from
the Sloan Digital Sky Survey. We found the observed halo boundary, the splachback
radius \({{R}_{{{\text{sp}}}}}\), which is equal to 4.12 Mpc
($M_r < -20\,.\!\!^{\rm m}3$) and and 4.06 Mpc ($M_r < -21\,.\!\!^{\rm m}5$)
over the integral
distribution of the number of galaxies as a function of the squared distance
from the center. We have studied how early-type galaxies are distributed
in the center and in the outskirts of the cluster
($R/R_{200}$<3, $M_r < -20\,.\!\!^{\rm m}3$)
and plotted the red sequence in the form of
\((g - r) = ( - 0.024 \pm 0.001){{M}_{r}} + (0.441 \pm 0.005)\).
Among all the cluster galaxies, the galaxies with the quenched star formation
(\( - 12\) yr\(^{{ - 1}} < \log sSFR < - 10.75\) yr\(^{{ - 1}}\)) make up about one
third. We have found that the fraction of QGs beyond the splashback
\({{R}_{{{\text{s}}p}}}\) is the same as in the field at the same \(z\) with
coordinates of the center of $16\,.\!\!^{\rm h}5$, $31^\circ$
and a size of $300'$. For galaxies with the stellar masses
\(\log {{{{M}_{*}}} \mathord{\left/ {\vphantom {{{{M}_{*}}} {{{M}_{ \odot }}}}} \right.
\kern-0em} {{{M}_{ \odot }}}} = [10.5;11.0]\) (this is the main mass range of QGs), after
entering the cluster, there is a decrease in the radii \({{R}_{{90,r}}}\) by about \(30\% \)
when moving towards the center.
\vspace{0.3cm} \noindent

\end{abstract}
\maketitle

\section{INTRODUCTION}
Galaxy clusters, the largest gravitationally bound systems in the Universe,
are the main objects of the observational cosmology \cite{Vikhlinin}
allowing one to study the hierarchical growth of structures. For such studies,
one of the best targets in the local the Universe (\(z < 0.1\)) is the cluster
A 2142 with the highest luminosity in the X-ray  and with a large number
of galaxy groups in its outskirts, which it gravitates. When observing the
cluster in the X-ray  (Chandra, XMM-Newton), multiple concentric cold
fronts of surface brightness \cite{Markevitch0, Rossetti}
were found indicating the sloshing activity of the core \cite{Tittley, Markevitch1}
stretching up to 1 Mpc from
the center of the cluster \cite{Rossetti}. A two-component radio
halo \cite{Venturi} was detected in the radio range (observations
with LOFAR and VLA). Eckert et al. (2014) also found evidence of accretion
of gas and galaxies onto the cluster: the galaxy group was discovered at
a distance of 1.5 Mpc northeast of the main system.

The central region of the A2142 has been studied in the optical range.
Owers et al. (2011) identified several subclusters falling on it in the
distribution of galaxies near the cluster (up to 2 Mpc). Einasto et al.(2020)
have shown that the cluster A2142 is the result of past and present
galaxy mergers. Liu et al. (2018) found 19 subclusters within 3.5 Mpc and
studied multiple mergers of smaller subsystems with the main cluster, which
influenced its complex dynamics. In addition, in the indicated paper, the
gas rings were found near two cold fronts seen in the X-ray region. The
velocities of the rings exceed the average radial velocity of the cluster
by \(810 \pm 330\) and \(660 \pm 300\) km s\(^{{ - 1}}\). Liu et al.(2018)
explained these features of the gas in terms of a core sloshing activity model.

The objectives of this paper are to determine the boundary of the A 2142
and the characteristics of its constituent galaxies: the fraction of early-type
galaxies in the red sequence and galaxies with the quenched star formation along the
normalized radius, as well as to reveal changes in the radius \({{R}_{{90,r}}}\)
of the galaxies along the radius of the cluster.

We carried out the study using the data from the SDSS catalog (Sloan Digital Sky
Survey Data Release 7 \cite{Abazajian}, 10 \cite{Aihara}), and NED
(NASA Extragalactic Database). The paper is organized as follows. Section 2
describes the procedure of measuring the dynamical characteristics of the galaxy
cluster and determining its surroundings. Section 3.1 presents the plotted
color-magnitude relation of early-type galaxies, estimates of the fraction of early-type
galaxies along the radius and their characteristics at a fixed stellar mass, and
compares this fraction with the values obtained for field objects. In Section 3.2,
the specific star formation rate of galaxies is considered; galaxies with the
quenched star formation (QGs) are found. The fraction of QGs along the radius
is determined at a fixed stellar mass of galaxies, their characteristics are
given, and a comparison is made with the specific star-formation rate in the
field. Section 4 lists the main results. In the paper, we used the following
cosmological parameters:
\({{\Omega }_{m}} = 0.3\), \({{\Omega }_{\Lambda }} = 0.7\) and
\({{H}_{0}} = 70\) km s\(^{{ - 1}}\) Mpc\(^{{ - 1}}\).

\section{DYNAMICAL CHARACTERISTICS}

The completeness of the galaxy cluster under study is determined by the
completeness of the spectroscopic data of the SDSS catalog. Strauss et al. (2002)
applied the following conditions when selecting the galaxies (the Main galaxy
sample): \({{r}_{{{\text{Pet}}}}} < 17\,.\!\!^{\rm m}77$ and
\(\langle {{\mu }_{r}}\rangle < 24\,.\!\!^{\rm m}5$,
where \({{r}_{{{\text{Pet}}}}}\) is the Petrosian magnitude of the galaxy in
the \(r\) filter corrected for extinction in the Galaxy, and \(\langle {{\mu }_{r}}\rangle \)
is the Petrosian average surface brightness corresponding to the
effective radius. Under these conditions, the completeness of the SDSS galaxies
was \(99\% \) and for bright galaxies \(95\% \).

Usually, when studying galaxy clusters, to increase the spectral completeness of bright galaxies
in the absence of the SDSS measurements of their radial velocities, we added the data from the NED
database. If there were no such measurements in the NED, we selected early-type bright red sequence
galaxies as probable cluster members (from one to five galaxies) using the
color-magnitude diagrams (\((u - r),{{M}_{r}}\); \((g - r),{{M}_{r}}\); and \((r - i),{{M}_{r}}\))
\cite{Kopylova18, Kopylova19}.

Determination of the dynamical characteristics of the system of galaxies: the radial velocity, the
dispersion of the radial velocities and mass is performed for a region of the radius \({{R}_{{{\text{2}}00c}}}\).
The radius \({{R}_{{{\text{2}}00c}}}\) (hereinafter \({{R}_{{{\text{2}}00}}}\)) is the empirical radius,
within which the density in the system 200 times exceeds the critical density of the Universe and can be
estimated by the formula:
\({{R}_{{200}}} = {{\sqrt 3 \sigma } \mathord{\left/ {\vphantom {{\sqrt 3 \sigma } {(10H(z))}}} \right. \kern-0em} {(10H(z))}}\) Mpc
\cite{Carlberg}. Then, assuming that the cluster is virialized within this radius, \({{M}_{{200}}} \sim {{M}_{{{\text{vir}}}}}\),
we find the mass \({{M}_{{200}}} = 3{{G}^{{ - 1}}}{{R}_{{200}}}\sigma _{{200}}^{2}\). That is, the mass of the cluster measured
by us \({{M}_{{200}}} \propto {{\sigma }^{3}}\). The mass \({{M}_{{200}}}\) enclosed in a region of the radius \({{R}_{{200}}}\)
can be determined directly from the critical density which depends on \(z\):

$${{M}_{{200}}} = \frac{4}{3}\pi R_{{200}}^{3} \times 200{{\rho }_{c}}.$$

In model calculations, the radius \({{R}_{{{\text{2}}00{\text{m}}}}}\) is often used; this is the radius, within which
the density in the system 200 times exceeds the average density of the Universe.
We present the main cluster parameters for the region with the radius \({{R}_{{200}}}\) (according to the SDSS DR7 data): the
radial velocity dispersion of galaxies, the radii (\({{R}_{c}}\), \({{R}_{{200}}}\), and \({{R}_{{{\text{sp}}}}}\)), the heliocentric
redshift, the dynamical mass, and the X-ray luminosity in the paper by Kopylova and Kopylov (2015) with the corresponding references and in the
columns of Table 1. Our estimates of \({{R}_{{200}}}\) and \({{M}_{{200}}}\) for the cluster are in good agreement with the data from
the paper by \cite{Tchernin} which collects the results of determining these parameters by the methods different from ours.

The structure and kinematics of A 2142, as well as its immediate surroundings, can be characterized in more detail using the
panels of Fig. 1: (a) the deviation of the radial velocities of galaxies--
the cluster members and background galaxies-- from the  average radial velocity depending on the square of the distance from
the cluster center; (b) the location of galaxies in the sky plane in equatorial coordinates; (c) the integral distribution of the
number of galaxies depending on the square of the distance from the center; (d) the histogram of the distribution of radial velocities of all the galaxies
within the radius \({{R}_{{200}}}\).
Note (see Fig. 1a) that the cluster is located along the line of sight, except for a group of galaxies with \(cz < 2500\) km s\(^{{1}}\)
at the bottom of the panel. Figure 1b shows the cluster profile in the projection (the integral distribution of the number of
galaxies depending on the squared radius from the center). It shows that the cluster first has a steep increase in the
number of galaxies, and then the number of galaxies located outside the virialized region grows linearly (the straight line
in the figure). In our paper \cite{Kopylov}, a cluster halo-bounding radius \({{R}_{h}}\) was defined, which we later
identified with the splashback radius. On closer examination, it turned out that for massive galaxy clusters of the A 1656 and
A 2142 types, it is necessary to use \(3\)-\(4\) radii \({{R}_{{200}}}\) to find the distribution point for the linear section.
In the cluster A 2142, the radius \({{R}_{{{\text{sp}}}}} = 4.12\) Mpc
(\({{M}_{r}} < - 20\,.\!\!^{\rm m}3$), \({{{{R}_{{{\text{sp}}}}}} \mathord{\left/ {\vphantom {{{{R}_{{{\text{sp}}}}}} {{{R}_{{{\text{200c}}}}}}}}
\right. \kern-0em} {{{R}_{{{\text{200c}}}}}}} = 1.81\) or \({{{{R}_{{{\text{sp}}}}}} \mathord{\left/ {\vphantom {{{{R}_{{{\text{sp}}}}}} {{{R}_{{{\text{200m}}}}}}}}
\right. \kern-0em} {{{R}_{{{\text{200m}}}}}}} = 1.13\). From brighter galaxies with \({{M}_{r}} < - 21\,.\!\!^{\rm m}5$), we found
\({{R}_{{{\text{sp}}}}} = 4.06\) Mpc. According to the results of the paper (see \cite{Kopylova22}, Fig. 5), the following relationship
holds for the A 2142 cluster:

$$\log {{R}_{{{\text{sp}}}}} = (0.24 \pm 0.03)\log {{L}_{X}} - (7.39 \pm 0.33).$$

The dotted line in Fig. 1 shows the radius of the virialized region $R_{200}$, the dashed line is the radius of the central
region $R_c$, the dash-dotted line corresponds to the splashback radius
$R_{sp}$ and is followed by a steep rise of the number of cluster members replaced by a linear one.
Figure 1 below also shows the distribution of bright early-type galaxies (at these \(z\), we estimated the limit \({{M}_{r}}\) for A~2142 approximately equal
to $-20\,.\!\!^{\rm m}3$ from which this radius is refined). The observed radius \({{R}_{{{\text{sp}}}}}\) (usually \({{R}_{{{\text{sp}}}}} > {{R}_{{200}}}\))
is the radius of the apocenter of the first orbits of galaxies, to which galaxies fly out of the virialized region after the first passage through
the cluster center. That is, the radius \({{R}_{{{\text{sp}}}}}\) separates most galaxies that fall into the cluster for the first time from the
collapsing galaxies that are already participating in the virial equilibration. The paper by \cite{Haines} (Fig. 13) shows the positions
of galaxies of all kinds, including those that have escaped from the cluster, in the phase space diagram obtained from model simulations of galaxy clusters.

We measured the radii \({{R}_{{{\text{sp}}}}}\) for a sample of 157 galaxy clusters. They vary in the range from 1.02\({{R}_{{200}}}\) to 3.64\({{R}_{{200}}}\)
depending on the dynamical mass and X-ray luminosity \cite{Kopylova22} and on average are equal to 1.54\({{R}_{{200}}}\).

\section{SAMPLES OF GALAXIES AND THEIR PROPERTIES}

\subsection{Early-type galaxies}

It is known that the main population of galaxy clusters at low redshifts (\(0 < z < 0.1\)) are early-type galaxies which are located mainly in
the central virialized regions and are the brightest members. Early-type galaxies follow the color-magnitude relationship called the
red sequence (RS).  In groups and clusters of galaxies, the RS has a small scatter, since the galaxies are at the same distance. We have found
that such galaxies in the virialized regions of galaxy clusters (for example, in the Hercules and Leo superclusters of galaxies) are of
the order of \(60\)-\(70\%\) among galaxies brighter than \({{M}_{K}} = - 23\,.\!\!^{\rm m}3$ \cite{Kopylova3}.
Early-type galaxies in this paper are selected according to the following criteria (the \(r\)-filter):
1) \(fracDeV \geqslant 0.8\), where the parameter \(fracDeV\) (according to the SDSS catalog) characterizes the contribution of the bulge to
the galaxy surface-brightness profile;
2)  the concentration index \(c \geqslant 2.6\), where \(c = {{{{r}_{{90}}}} \mathord{\left/ {\vphantom {{{{r}_{{90}}}} {{{r}_{{50}}}}}}
\right. \kern-0em} {{{r}_{{50}}}}}\) (is equal to the ratio of the radii limiting 90 and \(50\% \) of the Petrosian fluxes).
In addition, we made the \((u - r)\) color restrictions to exclude spiral galaxies: \(\Delta (u - r) \geqslant - 0.2\) which follows
from the dependence between the \((u - r)\) color and the absolute magnitude
$$(u - r) = - 0.108{{M}_{r}} + 0.63$$ with \(2\sigma = 0.2\). One can also make the \((g - r)\) color restrictions to narrow the
RS: \(\Delta \left| {(g - r)} \right| \geqslant 0.075\). We took the colors of the galaxies from the SDSS catalog calculated from
model magnitudes corrected for the extinction in the Galaxy.

The RS obtained for the A 2142 cluster is shown in Fig. 2a against the background of other galaxies. Figure 2b shows the
phase space diagram
of the A\,2142 in the projection, where \({{\Delta V} \mathord{\left/ {\vphantom {{\Delta V} \sigma }} \right. \kern-0em}
\sigma }\) is the ratio of the radial velocity difference of galaxies and the average radial velocity of the cluster to the
dispersion of radial velocities, \({R \mathord{\left/ {\vphantom {R {{{R}_{{200}}}}}} \right. \kern-0em} {{{R}_{{200}}}}}\)
is the distance of the galaxy from the selected cluster center normalized to the radius \({{R}_{{200}}}\). We took the brightest
galaxy as the center of the system of galaxies, whose coordinates were close to the coordinates of the X-ray center. The solid
circles show the RS galaxies within a radius of \({{3.5R} \mathord{\left/ {\vphantom {{3.5R}
{{{R}_{{200}}}}}} \right. \kern-0em} {{{R}_{{200}}}}}\) (Fig. 2a and Fig. 2b), the open circles mark the rest of the galaxies. The dashed
model line \cite{Barsanti, Oman}) roughly limits the virialized members of the cluster. We took the galaxies
with \( - 2 < {{\Delta V} \mathord{\left/ {\vphantom {{\Delta V} \sigma }} \right. \kern-0em} \sigma } < 2\) as the cluster
members beyond the radius \({{R}_{{200}}}\), therefore, the falling group galaxies (Fig. 1) with \(\Delta cz\) about \( - 2500\)
km s\(^{{ - 1}}\) is not shown in the phase diagram.

Some characteristics of early-type galaxies and estimates of their fraction \(fra{{c}_{{\text{E}}}}\) along the cluster radius
are given in Table 2. The first line contains the results for the sample of galaxies on the RS, restricted by the color
\(\Delta \left| {(g - r)} \right| < 0.075\), as in the papers by \cite{Kopylova18, Kopylova19} (for 40
clusters of galaxies with \(z < 0.045\). The second line gives the results without the \((g - r)\) restriction. In total,
we found 214 early-type galaxies on the RS in the cluster A 2142 (\({R \mathord{\left/ {\vphantom
{R {{{R}_{{200}}}}}} \right. \kern-0em} {{{R}_{{200}}}}} < 3\)). If we introduce the color \((g - r)\) restrictions,
we get 189 galaxies. It can be noted that the fractions of early-type galaxies on the RS vary along the radius of the
A 2142 cluster in the same way as, on average, for nearby galaxy clusters \cite{Kopylova19}. In order to
compare the results obtained for the very rich cluster A 2142 and the results obtained for low-density
regions near the cluster, we took a field of a size of $300'$ at the same redshifts (\(0.0815 < z < 0.0988\))
with the center coordinates ($16\,.\!\!^{\rm h}5$, $31^\circ$).
With the same selection of early-type galaxies (and taking into account the same condition $M_r$ < - $20\,.\!\!^{\rm m}3$
as for the cluster, we found that the fraction of early-type galaxies in the field is \(0.28 \pm 0.05\). This is
consistent with the values given by Table 2 for the region beyond
\({{R}_{{{\text{sp}}}}} = 1.81{{R}_{{200}}}\) or within 2-3
\({R \mathord{\left/ {\vphantom {R {{{R}_{{200}}}}}} \right. \kern-0em} {{{R}_{{200}}}}}\). This table also gives
the average characteristics of early-type galaxies for the main range of stellar masses
\(\log {{{{M}_{*}}} \mathord{\left/ {\vphantom {{{{M}_{*}}} {{{M}_{ \odot }}}}} \right.
\kern-0em} {{{M}_{ \odot }}}} = [10.5;11.0]\). It can be noted that the age of galaxies (in the SDSS DR10, it is
defined as the mass-weighted average age of the stellar population in Gyr) slightly decreases, the average
effective radius increases by 10\%, the average metallicity (\({{Z}_{ \odot }} = 0.019\)) of the stellar
population changes slightly outside the cluster radius \({{R}_{{200}}}\). It can also be noted that early-type
galaxies in the center of the cluster are redder, their effective radius is smaller, they are older and
richer in metals than galaxies within a radius of \(1\)-\(2{{R}_{{200}}}\). The characteristics of
galaxies outside the radius \({{R}_{{{\text{sp}}}}}\) differ from the values given for the field.
The galaxies in the field are slightly older, richer in metals, and have a smaller radius \({{R}_{e}}\).
The RS plotted from 214 galaxies is described by the expression
$$(g - r) = - 0.024( \pm 0.001){{M}_{r}} + 0.441( \pm 0.005)$$
with \({\text{rms}} = 0.032\). Within error, the shape and zero-point practically do not change with
the radius of the cluster.

\subsection{Galaxies with the quenched star formation}

The so-called~\glqq{Main Sequence}\grqq~relates the star formation rate of spiral galaxies to the stellar mass.
As a result of the depletion of gas in galaxies, the star-formation rate falls, and the galaxy
moves towards a passive state passing through an intermediate state characterized by the quenched
star formation. Within clusters of galaxies, there are different mechanisms that lead to the
quenching of star formation. In the central regions, these are tidal effects, as a result of
which galaxies are deprived of gas, stars, and dark matter (e.g., \cite{Mayer}). Some
galaxies lose matter, when they are in small groups of galaxies, even before they get into
the cluster; that is, they experience pre-processing (e.g., \cite{Poggianti, Wetzel, Haines1}).

In Fig. 1b, one can see that the rich cluster A 2142 is surrounded by many groups
of galaxies, in which, most likely, pre-processing of galaxies takes place. Earlier,
in the paper by \cite{Kopylova19}, for 40 nearby clusters of galaxies
(\(0.02 < z < 0.045\)), we found that even within the radius \(2 < {R \mathord{\left/
{\vphantom {R {{{R}_{{200}}}}}} \right. \kern-0em} {{{R}_{{200}}}}} < 3\)
(in the outskirts of the considered clusters) the fraction of the galaxies
with the quenched star formation is by 27\% larger than that in the field.

The specific star formation rate \(sSFR\) in the galaxy is defined as the integral
star formation rate divided by its stellar mass,
\(sSFR = {{SFR} \mathord{\left/ {\vphantom {{SFR} {{{M}_{*}}}}} \right. \kern-0em} {{{M}_{*}}}}\).
The SDSS DR10 catalog contains the results of determining the specific
star formation rate, the stellar mass of galaxies, and other parameters
galaxies, which have been obtained by fitting the FSPS \cite{Conroy}
models to the SDSS photometry in the \(u\), \(g\), \(r\), \(i\), and \(z\) filters.
We used the model values corrected for extinction and the \glqq{early-star formation with dust}\grqq~
version. In the distribution of galaxies according to the specific
star formation rate, \(\log sSFR\), a minimum is usually found that separates
galaxies with active star formation (active galaxies) from galaxies,
in which it is quenched (quenched galaxies--QGs) \cite{Wetzel1}.
In general, the distribution of galaxies in terms of the specific
star formation rate \(\log sSFR\) has a long tail extending into
the region of galaxies without star formation (passive galaxies).
In our papers \cite{Kopylova18, Kopylova19}, we selected QGs and
passive galaxies based on the condition
\(\log sSFR < - 10.75\) yr\(^{{ - 1}}\).
If we exclude galaxies without star formation from
the sample (\(\log sSFR < - 12\) yr\(^{{ - 1}}\) according
to Oemler et al. (2017), there will remain galaxies
with the quenched star formation, i.e., those satisfying
the condition \( - 12\) yr\(^{{ - 1}} < \log sSFR < - \)10.75 yr\(^{{ - 1}}\).

In the studied cluster A 2142 (\({R \mathord{\left/
{\vphantom {R {{{R}_{{200}}}}}} \right. \kern-0em} {{{R}_{{200}}}}} < 3\)),
we found 188 galaxies with the quenched star formation, 147 of
which are early-type galaxies with the parameter \(fracDeV \geqslant 0.8\),
the remaining 41 are late-type galaxies. Out of 188 galaxies,
120 have the stellar masses within
\(\log {{{{M}_{*}}} \mathord{\left/ {\vphantom {{{{M}_{*}}} {{{M}_{ \odot }}}}}
\right. \kern-0em} {{{M}_{ \odot }}}} = [10.5;11.0]\), 61 galaxies have
\(\log {{{{M}_{*}}} \mathord{\left/ {\vphantom {{{{M}_{*}}} {{{M}_{ \odot }}}}}
\right. \kern-0em} {{{M}_{ \odot }}}} = [11.0;11.5]\), and seven
more are not included in this range; that is, the galaxies
we study are massive (\({{\log {{M}_{*}}} \mathord{\left/
{\vphantom {{\log {{M}_{*}}} {{{M}_{ \odot }}}}} \right. \kern-0em} {{{M}_{ \odot }}}} > 10.5\)).

Figure 2c shows (with the solid circles) QGs against the background of
other galaxies. The first line of Table 3 shows the total fractions of
QGs and passive galaxies (similar to those given in the paper by \cite{Kopylova19},
the second line shows only the fractions of QGs along
the radius of the A\,2142 cluster. It can be noted that the fraction of QGs
(including passive ones) is maximum in the central region, somewhat
smaller within \({{R}_{{200}}}\), and drops by about 35\% beyond
\({{R}_{{{\text{sp}}}}} = 1.81{{R}_{{200}}}\). Beyond the splashback radius,
we have the same number of QGs as in the field \(0.58 \pm 0.10\) (285 galaxies
out of 491, $M_r <-20\,.\!\!^{\rm m}3$). At the same time, the fraction of QGs
without passive galaxies (the second line of Table 3) is minimum in the
center, maximum within the \({{R}_{{200}}}\) radius, and decreases
beyond the radius \({{R}_{{{\text{sp}}}}}\) by 28\% compared to those
in the virialized region of galaxy clusters (0-1\({{R}_{{200}}}\)).
We have found that the number of such galaxies in the field is
approximately the same: \(0.31 \pm 0.09\), 151 galaxies out of
491 ($M_r <-20\,.\!\!^{\rm m}3$).

Among QGs, the fraction of late-type galaxies
identified by the parameter \(fracDeV\) is
insignificant and amounts to 22\%. In the center
of the \(R < {{0.25R} \mathord{\left/ {\vphantom
{{0.25R} {{{R}_{{200}}}}}} \right. \kern-0em} {{{R}_{{200}}}}}\)
cluster, we found only three. Table 4 shows some properties of
late- and early-type QGs in the cluster and in the field at the
fixed stellar mass \(\log {{{{M}_{*}}} \mathord{\left/
{\vphantom {{{{M}_{*}}} {{{M}_{ \odot }}}}} \right. \kern-0em}
{{{M}_{ \odot }}}} = \) [10.5; 11.0]: the stellar population age,
metallicity, color \((g - r)\), concentration index \(c\),
and bulge fraction. The last column of the table shows the
parameters of the field galaxies (for \(52\) late-type galaxies
and \(48\) early-type galaxies). It can be noted that the
late-type galaxies falling into the cluster (within \({{R}_{{200}}}\)),
become redder, more compact, and similar to early-type galaxies:
the concentration index \(c\) increases from 2.39
beyond \({{R}_{{{\text{sp}}}}}\) to 2.56, the fraction of
the bulge \(fracDeV\) increases significantly from 0.36 to
0.56 (or by 36\%). For early-type galaxies, no significant
changes in the parameters are observed. If we compare these
parameters of QGs (for the same range of stellar masses) with
those from a lower density region (of the field), we can
draw the following conclusion. In the cluster region (\(R <
{{R}_{{{\text{sp}}}}}\)), late-type QGs have a larger bulge,
a redder color, and significantly lower metallicity
compared to the field galaxies.

In the paper \cite{Kopylova20}, we
considered the variations in the stacked Petrosian
radius of galaxies in 40 nearby (\(0.02 < z < 0.045\))
galaxy clusters. For the mass range

$$\log {{{{M}_{*}}} \mathord{\left/ {\vphantom
{{{{M}_{*}}} {{{M}_{ \odot }}}}} \right. \kern-0em}
{{{M}_{ \odot }}}} = [10.5;11.0],$$

it turned out that the radius \({{R}_{{90,r}}}\) of
late-type galaxies, when moving inside clusters,
decreases by about \(13\% \), and for early-type galaxies--by \(11\% \).
In this paper, we also studied the change in the average Petrosian radius
\({{R}_{{90,r}}}\) of the galaxies from the A 2142 cluster along its
radius for the same range of stellar masses.

Figure 3 shows the change in the average radius \({{R}_{{90,r}}}\) of the
galaxies with the quenched star formation, the members of the cluster
A 2142, along the normalized radius \({R \mathord{\left/ {\vphantom
{R {{{R}_{{200}}}}}} \right. \kern-0em} {{{R}_{{200}}}}}\). The solid
line corresponds to the late-type galaxies (\(fracDeV < 0.8\)), the
dashed line corresponds to the early-type galaxies (\(fracDeV \geqslant 0.8\)).
The same horizontal lines show the values for field galaxies. We can draw
the following conclusions. There are few QGs in the central region of
A 2142, but they do exist. Perhaps, this is the result of a projection
of galaxies that do not belong to the cluster. For the galaxies of late
and early types that fall into the virialized region of the cluster
(\(R < {R \mathord{\left/ {\vphantom {R {{{R}_{{200}}}}}} \right. \kern-0em} {{{R}_{{200}}}}}\)),
the radii \({{R}_{{90,r}}}\) gradually decrease towards the center by approximately 30\%.
Moreover, one can notice (Fig. 3) that the early-type galaxies have the maximum
radius \({{R}_{{90,r}}}\) near the splashback radius,
this is not observed for the late-type galaxies. Thus, in the rich cluster
A 2142, all the galaxies with the quenched star formation with the mass
\(\log {{{{M}_{*}}} \mathord{\left/ {\vphantom {{{{M}_{*}}} {{{M}_{ \odot }}}}} \right.
\kern-0em} {{{M}_{ \odot }}}}\) [10.5;11.0] show significant decreases of the radii.

Since the cluster A 2142 is very massive and many groups of galaxies and individual
galaxies fall on it even near a radius of 3.5\({R \mathord{\left/ {\vphantom {R {{{R}_{{200}}}}}}
\right. \kern-0em} {{{R}_{{200}}}}}\) (which can be seen in Fig. 1), the QGs
have radii of \({{R}_{{90,r}}}\) somewhat smaller than those in the field,
although, within error. Also, we do not detect the effect we found in the
paper \cite{Kopylova20} for nearby clusters of galaxies, when
the stacked radii \({{R}_{{90,r}}}\) of the late-type galaxies are maximum near
the average stacked splashback radius of clusters.
Apparently, the reason is that in A 2142 we deal only with massive galaxies
(\({{M}_{r}} < - 20\,.\!\!^{\rm m}3$); while in the paper by Kopylova and
Kopylov (2020), we studied mostly fainter and less massive galaxies with
\({{M}_{r}} > - 20\,.\!\!^{\rm m}3$ in nearby clusters. They are 1.5
times greater in number than massive galaxies. That is, the effect we found
(although this was not emphasized in the paper by \cite{Kopylova20})
appears mainly in smaller galaxies, since they are easier to destroy in the
field of galaxy clusters. It is known that clustering of galaxies and gas
occurs near the radius \({{R}_{{{\text{sp}}}}}\) \cite{Adhikari}
which results in acceleration of the rate of galaxy changes.

Galaxy clusters can be considered as laboratories, where transformations of
galaxies take place. In the papers \cite{Poggianti1, Cebrian}
it is shown that early-type galaxies in clusters are smaller
than those in the field. Matharu et al.(2019), for example, found that the
increase in the size of passive and star-forming galaxies (\(z \sim \)1) in
the field compared to those in clusters can be explained by the collision and
merging of galaxies. Hamabata et al. (2019)  found the radius of the cluster (\(r\)
is about \(0.2{{h}^{{ - 1}}}\) Mpc, where \(h\) = 0.7), within which the
transformations of spiral galaxies are most effective.
The paper by \cite{Pranger} shows how the environment affects
spiral galaxies. For a sample, of 700 galaxies (\(z < 0.063\), the SDSS
data), it was found that sizes of spiral galaxies, in terms of the radius
\({{R}_{e}}\) as a parameter, are by 15\% smaller in clusters than those
in the field, and the \(Sersic\) parameter is by 15\% smaller. Also,
the color \((g - r)\) of galaxies are redder in clusters.

\section{CONCLUSION AND FINDINGS}

According to the SDSS data, for the galaxy cluster
A 2142 which is the brightest in the X-ray of the
local Universe, we have considered the changes:
(a) the fraction of early-type galaxies along the radius
(in the projection); (b) the fraction of galaxies with
the quenched star formation within the nearest outskirts
(up to \({{3R} \mathord{\left/ {\vphantom {{3R} {{{R}_{{200}}}}}}
\right. \kern-0em} {{{R}_{{200}}}}}\)) in comparison
with the data for the field. The paper by \cite{Kopylova20}
showed the changes in the fraction of QGs depending
on the stellar mass of galaxies. Massive early-type galaxies
tend to be located in the center of the galaxy cluster.
The QGs sample that we
compiled has stellar masses in the range of \(\log {{{{M}_{*}}}
\mathord{\left/ {\vphantom {{{{M}_{*}}} {{{M}_{ \odot }}}}}
\right. \kern-0em} {{{M}_{ \odot }}}} = [10.5;11.5]\) (64\%
of them are in the range of \(\log {{{{M}_{*}}} \mathord{\left/
{\vphantom {{{{M}_{*}}} {{{M}_{ \odot }}}}} \right. \kern-0em}
{{{M}_{ \odot }}}} = [10.5;11.5]\)), that is, these are massive
and bright galaxies. We present some properties of
these galaxies obtained in that paper.

The key results of this study are the following:

1. From the observed cluster profile--the integral distribution
of the number of galaxies depending on the square of the distance
from the center--we have found the splashback radius
\({{R}_{{{\text{sp}}}}}\) which coincides with the apocenters of
the orbits of most galaxies that have already visited the center
of the system. The radius \({{R}_{{{\text{sp}}}}}\) of the cluster
A 2142 is \(4.12\) Mpc, \({{{{R}_{{{\text{sp}}}}}} \mathord{\left/
{\vphantom {{{{R}_{{{\text{sp}}}}}} {{{R}_{{{\text{200c}}}}}}}}
\right. \kern-0em} {{{R}_{{{\text{200c}}}}}}} = 1.81\) for the
critical density, or \({{{{R}_{{{\text{sp}}}}}} \mathord{\left/
{\vphantom {{{{R}_{{{\text{sp}}}}}} {{{R}_{{{\text{200m}}}}}}}}
\right. \kern-0em} {{{R}_{{{\text{200m}}}}}}} = 1.13\) for the
average density of the Universe.

2. The shape of the color-magnitude (the red sequence, RS) dependence
does not change within the considered cluster radius ranges (Table 1).
We have obtained the RS as
$$(g - r) = - 0.024( - 0.023){{M}_{r}} + 0.441(0.442).$$

The coefficients of the form and zero-point  are given for the region
with the radius \({{R}_{{{\text{sp}}}}}\) and beyond it (in parentheses).
The fraction of early-type galaxies outside \({{R}_{{{\text{sp}}}}}\)
is equal to the value for the low-density region, for which the field
with the coordinates $16\,.\!\!^{\rm h}5$, $31^\circ$ and the radius $300'$
at the same \(z\) is taken. The early-type galaxies in the center of
the cluster are redder, have a smaller effective radius \({{R}_{e}}\),
are older, and metal-richer than those outside the virial radius,
1--2$R_{200}$ (Table 1).

3. The fraction of galaxies with the quenched star formation
(\( - 12\) yr\(^{{ - 1}} < \log sSFR < - 10.75\) yr\(^{{ - 1}}\))
decreases by 28\% beyond the radius \({{R}_{{{\text{sp}}}}}\)
compared to the radius \({{R}_{{200}}}\) and becomes the same
as in the field. Among the QGs, only 22\% are late-type galaxies.
We have found that in such galaxies both the concentration index
and the fraction of the bulge increase towards the center of
the cluster, i.e., they become similar to early-type galaxies.

4. It is found that the properties of galaxies with the quenched
star formation with the stellar mass \(\log {{{{M}_{*}}} \mathord{\left/
{\vphantom {{{{M}_{*}}} {{{M}_{ \odot }}}}} \right. \kern-0em}
{{{M}_{ \odot }}}} = \) [10.5;11.0], when falling into the
rich massive cluster A 2142, change greatly. The radii \({{R}_{{90,r}}}\)
of the galaxies of late and early types decrease by about 30\%
in the center of the cluster compared to the radii near the
boundary of the cluster halo.

\begin{acknowledgments}
This research has made use of the NASA/IPAC Extragalactic Database
(NED, \url{http://nedwww.}\linebreak\url{.ipac.caltech.edu}),
which is operated by the Jet Propulsion Laboratory, California Institute of
Technology, under contract with the National Aeronautics and Space
Administration, Sloan Digital Sky Survey (SDSS, \url{http://www.sdss.org}),
which is supported by Alfred P. Sloan Foundation, the participant institutes
of the SDSS collaboration, National Science Foundation, and the United
States Department of Energy and Two Micron All Sky Survey (2MASS,
\url{http://www.ipac.}\linebreak\url{.caltech.edu/2mass/releases/allsky/}).
\end{acknowledgments}

\section*{CONFLICT OF INTEREST}
The authors declare no conflict of interest regarding this paper.

\begin{center}
\refname
\end{center}

\onecolumngrid
\newpage

\begin{table*}[p]
\setcaptionmargin{0mm} \onelinecaptionstrue
\captionstyle{flushleft}
\caption{Dynamical characteristics of  A\,2142}
\label{data1}
\medskip
\small
\begin{tabular}{c|c|c|c|c|c|c|c|c}
\hline
Cluster & $\sigma$ & $R_c$  & $R_{200}$ &  $R_{sp}$ & $N_z$& $z_h$ &  $M_{200}$ & $L_{0.1-2.4 keV}$ \\
\hline
	& km~$c^{-1}$& Mpc  &Mpc & Mpc &  &  & $10^{14}~M_{\odot}$ & $10^{44}$ erg $s^{-1}$ \\
\hline
(1) & (2) & (3) & (4) & (5) & (6) & (7) & (8) & (9)\\
\hline
A\,2142  & $963\pm70$ & 1.52 & 2.28 & 4.12  & 191 & 0.090135 & $14.82\pm3.23$ & 10.58    \\
\hline
\end{tabular}
\end{table*}

\begin{table*}[p]
\setcaptionmargin{0mm} \onelinecaptionstrue
\captionstyle{flushleft}
\caption{A 2142: early-type galaxies on the RS and their properties along the radius
}
\label{data2}
\medskip
\small
\begin{tabular}{l|r|r|r|r|r|r|r}
\hline
properties  & 0--0.25$R_{200}$ & 0--1$R_{200}$ & 1--2$R_{200}$ &  2--3$R_{200}$ & 0--1$R_{sp}$ & 1$R_{sp}$--3$R_{200}$ & field\\
\hline
(1) & (2) & (3) & (4) & (5) & (6) & (7) & (8)\\
\hline
$frac_{E1}$        & $0.51\pm0.14$ &$0.50\pm0.06$  & $0.35\pm0.05$      & $0.27\pm0.05$ & $0.44\pm0.04$ & $0.26\pm0.05$             & $0.27\pm0.03$ \\
$frac_{E2}$        & $0.78\pm0.19$ &$0.61\pm0.07$  & $0.36\pm0.05$      & $0.29\pm0.06$ & $0.51\pm0.05$ & $0.28\pm0.05$             & $0.30\pm0.04$ \\
$Age, Gyr$         & $9.48\pm0.20$ &$9.38\pm0.07$  & $9.32\pm0.09$      & $9.60\pm0.16$ & $9.35\pm0.06$ & $9.60\pm0.14$             & $9.94\pm0.11$\\
$M_r$              & $-20.76\pm0.10$ & $-20.79\pm0.04$ & $-20.82\pm0.04$& $-20.78\pm0.09$ & $-20.81\pm0.03$ & $-20.78\pm0.08$       & $-20.71\pm0.03$ \\
$\log(Z/Z_{\odot}$)& $-0.26\pm0.02$ &$-0.24\pm0.01$  & $-0.28\pm0.01$   & $-0.31\pm0.01$ & $-0.26\pm0.01$ & $-0.31\pm0.01$          & $-0.25\pm0.02$  \\
$(g-r$)            & $0.96\pm0.01$ &$0.94\pm0.01$  & $0.92\pm0.01$      & $0.92\pm0.01$ & $0.93\pm0.01$ & $0.92\pm0.01$             & $0.92\pm0.01$  \\
$R_e$, kpc         & $2.73\pm0.40$ &$2.84\pm0.14$  & $3.03\pm0.17$      & $2.88\pm0.25$ & $2.92\pm0.11$ & $2.84\pm0.23$             & $2.58\pm0.12$  \\
\hline
\end{tabular}
\end{table*}

\begin{table*}[p]
\setcaptionmargin{0mm} \onelinecaptionstrue
\captionstyle{flushleft}
\caption{A 2142: fraction of galaxies with the quenched star formation along the radius
}
\label{data3}
\medskip
\small
\begin{tabular}{l|c|c|c|c|c|c|c}
\hline
fraction  & 0--0.25$R_{200}$ & 0--1$R_{200}$ & 1--2$R_{200}$ &  2--3$R_{200}$ & 0--1$R_{sp}$ & 1$R_{sp}$--3$R_{200}$ & field \\
\hline
(1) & (2) & (3) & (4) & (5) & (6) & (7) & (8)\\
\hline
$frac_{q1}$ & $0.92\pm0.22$     &$0.85\pm0.09$     & $0.65\pm0.08$     & $0.63\pm0.09$     & $0.78\pm0.06$     & $0.60\pm0.08$  & $0.58\pm0.10$ \\
$frac_{q2}$ & $0.30\pm0.10$     &$0.42\pm0.06$     & $0.39\pm0.06$     & $0.32\pm0.06$     & $0.42\pm0.04$     & $0.30\pm0.05$  & $0.31\pm0.09$ \\
\hline
\end{tabular}
\end{table*}

\begin{table*}[p]
\setcaptionmargin{0mm} \onelinecaptionstrue
\captionstyle{flushleft}
\caption{A 2142: parameters of galaxies with the quenched star formation along the radius
($\log M^*/M_{\odot}$ = [10.5,11.0])
}
\label{data4}
\medskip
\small
\begin{tabular}{l|r|r|r|r|r|r|r}
\hline
properties  & 0--0.25$R_{200}$ & 0--1$R_{200}$ & 1--2$R_{200}$ &  2--3$R_{200}$ & 0--1$R_{sp}$ & 1$R_{sp}$--3$R_{200}$& field\\
\hline
(1) & (2) & (3) & (4) & (5) & (6) & (7) & (8)\\
\hline
$c$ ($R_{90,r}/R_{50,r}$)   & $2.65\pm0.01$   &$2.56\pm0.07$   & $2.53\pm0.06$   & $2.39\pm0.07$   & $2.54\pm0.04$   & $2.39\pm0.07$ & $2.45\pm0.03$ \\
$\it{frac}DeV$              & $0.78\pm0.01$   &$0.56\pm0.08$   & $0.55\pm0.03$   & $0.36\pm0.08$   & $0.55\pm0.04$   & $0.36\pm0.08$ & $0.41\pm0.04$ \\
$(g-r)$                     & $0.94\pm0.01$   &$0.89\pm0.01$   & $0.86\pm0.01$   & $0.89\pm0.03$   & $0.88\pm0.01$   & $0.89\pm0.03$ & $0.80\pm0.01$ \\
$age, Gyr$                  & $8.88\pm0.28$   &$8.66\pm0.08$   & $8.81\pm0.14$   & $8.70\pm0.12$   & $8.74\pm0.08$   & $8.70\pm0.12$ & $8.91\pm0.06$ \\
$\log(Z/Z_{\odot})$         & $-0.28\pm0.01$  &$-0.31\pm0.02$  & $-0.32\pm0.03$  & $-0.23\pm0.09$  & $-0.32\pm0.02$  & $-0.23\pm0.09$& $-0.08\pm0.03$ \\
\hline
$c$ ($R_{90,r}/R_{50,r}$)   & $2.69\pm0.03$   &$2.95\pm0.07$   & $2.86\pm0.04$   & $2.85\pm0.06$   & $2.90\pm0.04$   & $2.86\pm0.05$ & $2.86\pm0.02$ \\
$\it{frac}DeV$              & $0.90\pm0.04$   &$0.95\pm0.01$   & $0.94\pm0.01$   & $0.97\pm0.01$   & $0.95\pm0.01$   & $0.97\pm0.01$ & $0.96\pm0.01$ \\
$(g-r)$                     & $0.94\pm0.03$   &$0.91\pm0.01$   & $0.89\pm0.01$   & $0.90\pm0.01$   & $0.90\pm0.01$   & $0.89\pm0.01$ & $0.85\pm0.01$ \\
$age, Gyr$                  & $8.87\pm0.16$   &$8.92\pm0.05$   & $8.85\pm0.05$   & $8.90\pm0.06$   & $8.89\pm0.04$   & $8.89\pm0.05$ & $9.02\pm0.05$ \\
$\log(Z/Z_{\odot})$         & $-0.27\pm0.04$  &$-0.25\pm0.02$  & $-0.29\pm0.01$  & $-0.29\pm0.02$  & $-0.27\pm0.01$  & $-0.29\pm0.02$& $-0.24\pm0.02$ \\
\hline
\end{tabular}
\end{table*}

\begin{figure*}[b]
\setcaptionmargin{0mm}
\onelinecaptionsfalse
\includegraphics[scale=0.53,angle=0]{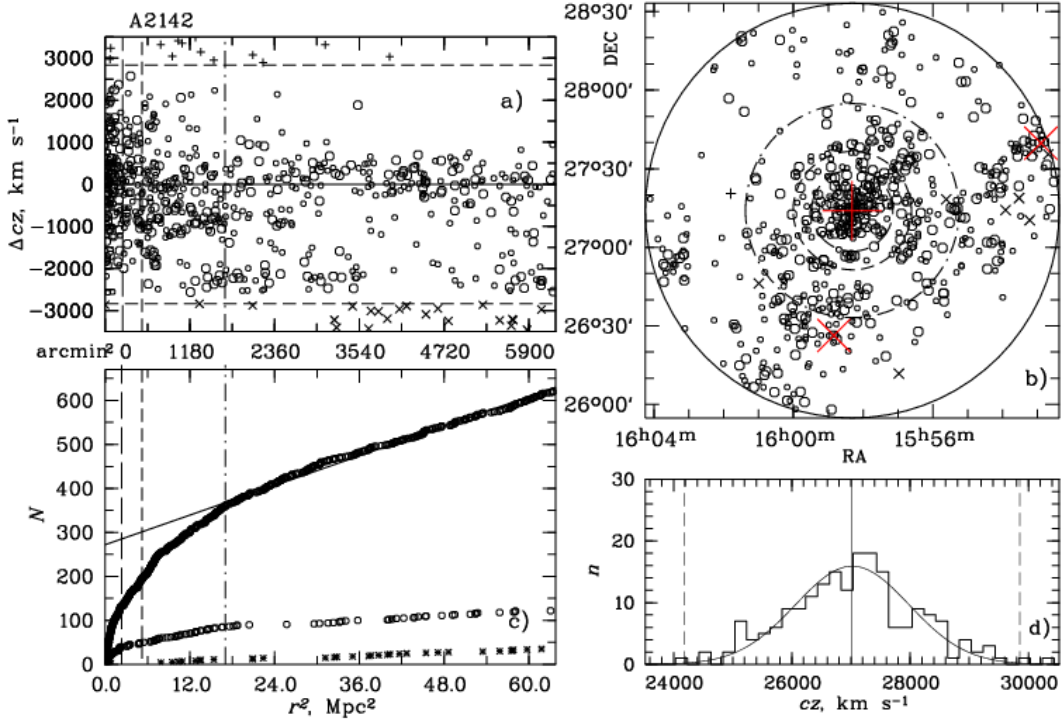}
\captionstyle{normal}
\caption{
Distribution of galaxies in the cluster A 2142: (a) deviation of the radial
velocities of galaxies from the average radial velocity of the cluster determined
from the galaxies within the radius $R_{200}$. The horizontal dashed lines
correspond to the deviations $\pm2.7\sigma$, the vertical lines mark
the radii $R_{200}$ (short dashes), $R_c$ (long dashes), $R_{sp}$ (dash-dotted).
Larger circles indicate galaxies brighter than $M_K^*+1 = -24^m$,
the plus marks are for background galaxies, and the crosses are foreground
galaxies. b) The distribution of the galaxies shown in panel (a) in the sky in
equatorial coordinate (the notation is the same). The areas with the radii $R_c$,
$R_{200}$ and $R_{sp}$ are marked with the circles. The research area is limited by a circle with a radius
of 3.5$R_{200}$ (the solid line). The large plus and crosses indicate
the center of the cluster (the brightest galaxy BCG1 with $M_r < -23\,.\!\!^{\rm m}82$),
and two other brightest galaxies (BCG2 with $M_r < -23\,.\!\!^{\rm m}49$,
BCG3 c $M_r < -23\,.\!\!^{\rm m}27$).
(c) The integral distribution of the total number of galaxies (the upper
curve) depending on the square of the distance from the center of the A2142. The lower curve
corresponds to the early-type RS galaxies (-0.075 < $(g-r)$ < 0.075) brighter than $M_r<-20^m.3$.
The circles correspond to the galaxies indicated by he circles in panel (a), the asterisks
correspond to the background galaxies. (d) The distribution by radial velocities of
all the galaxies within the radius $R_{200}$ (the solid line for the cluster members
shows the Gaussian corresponding to the radial velocity dispersion of the cluster).
The solid vertical line indicates the average radial velocity of the cluster, the
dashed lines correspond to the deviations $\pm2.7\sigma$.
}
\label{clus2142}
\end{figure*}

\newpage
\begin{figure*}[p]
\setcaptionmargin{5mm}
\onelinecaptionsfalse
\includegraphics[scale=0.58,angle=0]{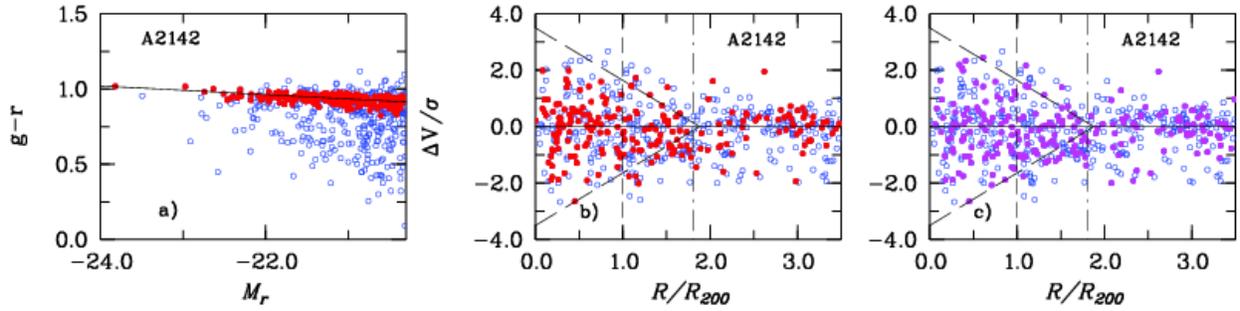}
\captionstyle{normal}
\caption{
(a) Color--absolute magnitude dependence for the galaxy cluster
A 2142. The RS galaxies (-0.075 < $(g-r)$ < 0.075) are shown with the solid red circles.
(b) and (c) the velocity--radius phase space diagrams, where velocity
is the ratio of the radial velocity difference of the galaxies and the average radial
velocity of the cluster to the radial velocity dispersion, and $R/R_{200}$
is the distance of the galaxy from the center of the cluster normalized to the
radius $R_{200}$. The dashed and dash-dotted lines show the radii $R_{200}$
and $R_{sp}$, respectively. The solid red circles in panel (b) are the RS galaxies,
as well as in panel (a). In panel (c), the solid magenta circles correspond to the galaxies
with the quenched star formation ($-3~Gyr^{-1} < \log sSFR < -1.75~Gyr^{-1}$).
The sloping dashed lines show the model calculations from \cite{Barsanti};
they separate the virialized cluster members.
}
\label{3figs}
\end{figure*}

\newpage
\begin{figure*}[p]
\setcaptionmargin{5mm}
\onelinecaptionsfalse
\includegraphics[scale=0.4,angle=0]{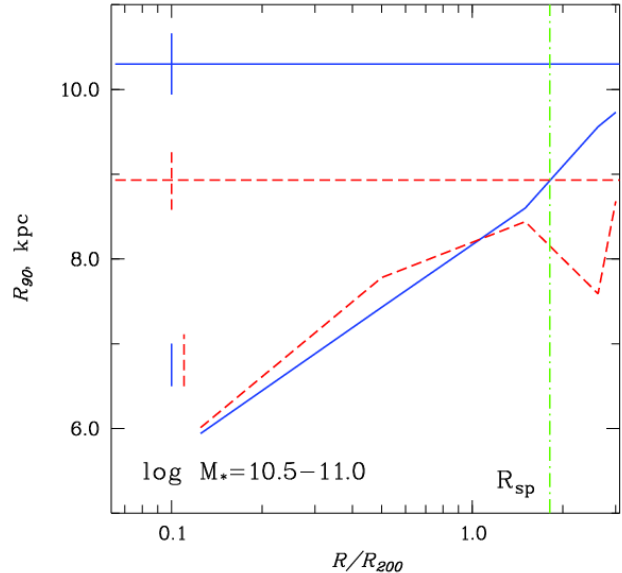}
\captionstyle{normal}
\caption{
Petrosian average radius $R_{90,r}$ in kiloparsecs of the galaxies
with the masses $\log M^*/M_{\odot}$ = [10.5,11.5] depending on the
normalized radius $R/R_{200}$. The dashed polygonal red line corresponds
to early-type galaxies $\it{frac}DeV \geq 0.8$, the solid blue line
corresponds to late-type galaxies with $\it{frac}DeV < 0.8$. The
average values of $R_{90,r}$ obtained from field galaxies of early and
late types are marked with horizontal lines of the same type.
The vertical short lines show the average errors in the measurements
of the radii. The vertical green line corresponds to the cluster radius $R_{sp}$.
}
\label{R90K}
\end{figure*}
\end{document}